\documentclass[letterpaper,english,showpacs,pre,preprint]{revtex4}
\usepackage[T1]{fontenc}
\usepackage[latin1]{inputenc}
\usepackage{color}
\usepackage{amsmath}
\usepackage{graphicx}
\usepackage{amssymb}

\makeatletter
\@ifundefined{textcolor}{}
{%
 \definecolor{BLACK}{gray}{0}
 \definecolor{WHITE}{gray}{1}
 \definecolor{RED}{rgb}{1,0,0}
 \definecolor{GREEN}{rgb}{0,1,0}
 \definecolor{BLUE}{rgb}{0,0,1}
 \definecolor{CYAN}{cmyk}{1,0,0,0}
 \definecolor{MAGENTA}{cmyk}{0,1,0,0}
 \definecolor{YELLOW}{cmyk}{0,0,1,0}
 }

\usepackage{graphicx}

\usepackage{graphicx}

\makeatother

\usepackage{babel}

\begin{document}

\title{Tunneling of Micro-sized Droplets Through a Flowing Soap Film}

\author{Ildoo Kim}

\affiliation{Department of Physics and Astronomy, University of Pittsburgh, Pittsburgh,
PA 15260}

\author{X.L. Wu}

\affiliation{Department of Physics and Astronomy, University of Pittsburgh, Pittsburgh,
PA 15260}

\date{March 26, 2010}
\begin{abstract}
When a micron-sized water droplet impacts on a freely suspended soap
film with speed $v_{i}$, there exists a critical impact velocity
of penetration $v_{C}$. For the droplet with $v_{i}<v_{C}$, it flows
with the soap film after the impact whereas with $v_{i}>v_{C}$, it
tunnels through. In all cases, the film remains intact despite the
fact that the droplet radius ($R_{0}=26\,\text{\ensuremath{\mu}m}$)
is much greater than the film thickness ($0<h\lesssim10\,\text{\ensuremath{\mu}m}$).
The critical velocity $v_{C}$ was measured as a function of $h$,
and interestingly $v_{C}$ approaches an asymptotic value $v_{C0}\simeq520\,${\normalsize cm/s}
in the limit $h\rightarrow0$. This indicates that in addition to
an inertial effect, a deformation or stretching energy of the film
is required for penetration. Quantitatively, we found that this deformation
energy corresponds to the creation of $\sim14$ times of the cross-sectional
area of the droplet ($14\pi R_{0}^{2}$) or a critical Weber number
$\text{W\ensuremath{e_{C}}}(\equiv2R_{0}\rho_{w}v_{C0}^{2}/\sigma)\simeq44$,
where $\rho_{w}$ and $\sigma$ are respectively the density and the
surface tension of water. 
\end{abstract}


\maketitle

\section{Introduction}

\textcolor{black}{When a liquid droplet impinges on a surface made
of the same liquid, a variety of phenomena occurs. If the underlying
liquid has a finite but large depth, the droplet causes splashing,
creating a crown-like structure that had become an iconic piece of
Edgerton's then newly invented strobe photography in the late 30's
\cite{Edgerton}. If the underlying liquid has a small thickness,
the macroscopic corona structure disappears altogether and instead
a thin ejecta sheet is created \cite{Thoroddsen2002}. The existence
of such ejecta sheet has only recently been discovered and has captured
the attention of scientists \cite{Josserand2003}. Underlying this
seemingly simple phenomenon is complex dynamics \cite{Xu2005,Josserand2003,Weiss1999},
whose understanding is currently incomplete but is fundamentally important
for a variety of industrial processes such as containment of hazardous
liquids, uniform coating of surfaces, and efficient fuel injection. }

\textcolor{black}{In this paper, we report a related subject, namely
the impact of a water droplet against a freely flowing soap film.
Specifically, we are interested in the condition of tunneling of a
ballistic droplet through the film. A classical analysis of Taylor
and Michael \cite{Taylor&Michael_JFM_1973} based on an energy argument
suggests that when a hole is created on a liquid film, it will shrink
and heal if its diameter is smaller than the film thickness. On the
other hand, if the diameter is larger than the thickness, an instability
that leads to the rupture of the film will occur. Analyses were also
carried out by Zheng and Witten \cite{Zheng&Witten_2006} in a proposal
to create a giant liquid film in space that is free of gravitational
forces and surrounding air; both are significant factors complicating
the study of two-dimensional fluid flows and turbulence using these
films in an earth environment \cite{Couder1984,Martin_PRL_1998,Vorobieff_PhysFluids_1999}.
Their calculation shows that a meteor of a few nanometer in size can
be hazardous to a space-based film. While the energetic argument is
compelling and has found some experimental confirmations \cite{Liang1996},
the calculation cannot account for certain observations made in these
films. For instance, common experiences show that when a soap film
is perturbed by a foreign object, the film often breaks. However,
if the object is wetted by water, the film is much more resistant
to the perturbation. A recent experiment demonstrated that a macroscopic
object as large as a tennis ball can readily pass through a micron-thick
film without breaking it \cite{Courbin&Stone_PhysFluid_2006}. It
is evident that the passage dynamics in this latter case cannot be
understood unless film deformation into the third dimension, perpendicular
to the film, is taken into account.}

\textcolor{black}{To investigate the tunneling process quantitatively
we implemented an inkjet technique to generate uniform sized ($R_{0}=26\,\text{\ensuremath{\mu}m}$)
water droplets at a controllable rate. The trajectories of the droplets
before and after impacts are digitized using a high-speed video camera,
allowing quantitative measurements of momentum and energy transfers
between the droplets and the film. It was observed that droplets can
tunnel through the soap film if its impact velocity is higher than
a certain threshold value $v_{C}$, and in no case rupture occurs
as a result of the impact. Using films of different thicknesses ($0<h\lesssim10\,\text{\ensuremath{\mu}m}$),
we found that the energy barrier $E_{C}=\frac{1}{2}mv_{C}^{2}$ for
tunneling is a linear function of $h$ and can be expressed as $E_{C}=E_{min}(1+\alpha h/R_{0})$,
where $E_{min}\simeq0.01\,\text{erg}$ and $\alpha\simeq3.9$ are
constants. Kinematically, we found that the tunneling process can
be modeled as an inelastic collision between the droplet and the film.
It only requires two parameters, $M_{1}$ and $M_{2}$, which specify
respectively the effective mass of the film involved in the collision
and the mass that is transferred to the droplet after the collision.}

\section{Experimental Setup}

\begin{figure}
\includegraphics[width=5cm]{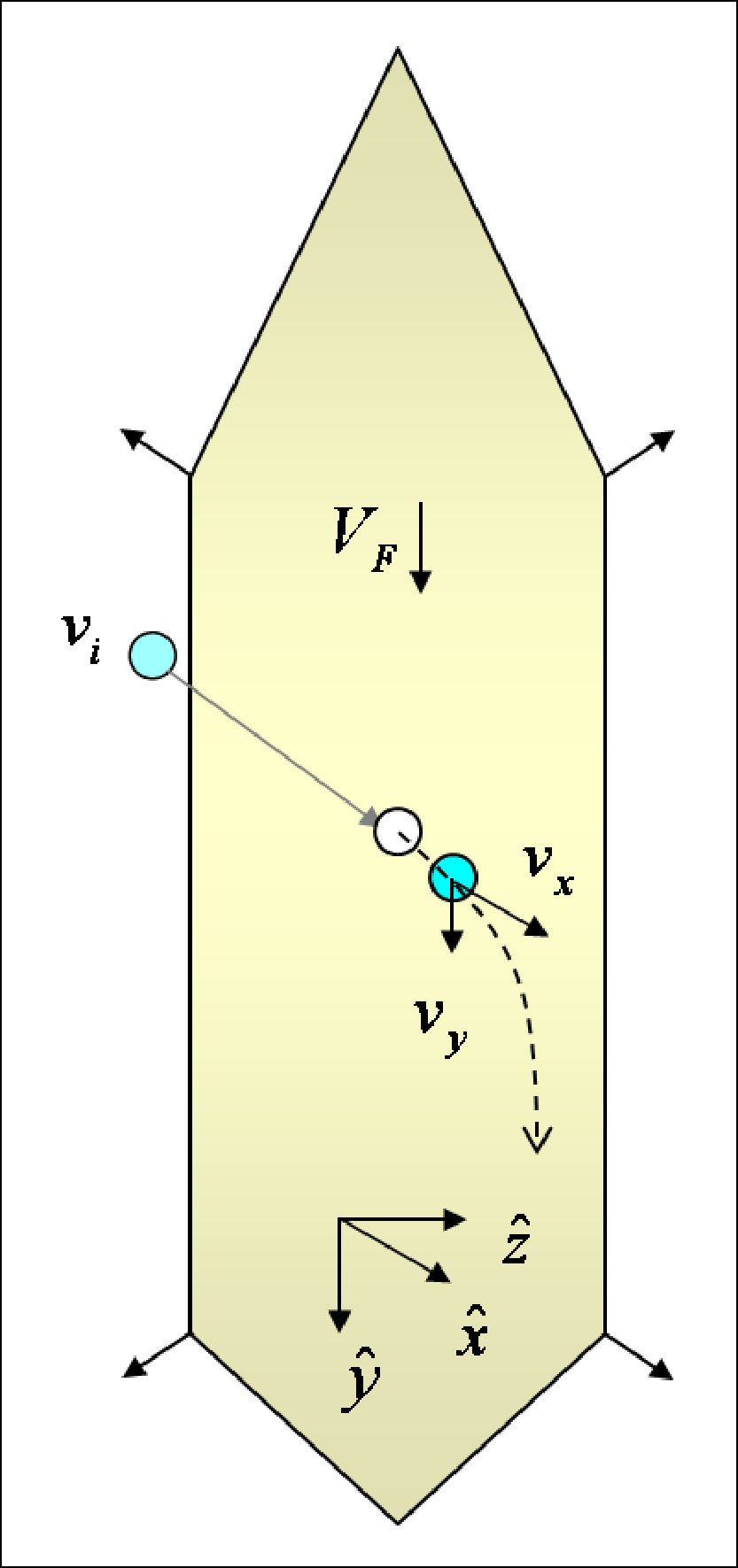}

\caption{The experimental setup. A vertical soap film flows in the $\hat{y}$
direction with a uniform velocity $V_{F}$. A water droplet is ejected
horizontally, in the $\hat{x}$ direction, toward the soap film with
an impact velocity $v_{i}$. The final velocity of the droplet after
impact $(v_{x},v_{y})$ is also indicated. To visualize the water
droplet and the film, a high-intensity halogen lamp and a low-pressure
sodium lamp were used. The droplet's trajectories and their corresponding
velocities were recorded by a high-speed video camera. \label{cap:expsetup}}
 
\end{figure}

Our experiment was carried out in a vertically flowing soap film as
depicted in Fig. \ref{cap:expsetup}. The construction of the soap-film
apparatus has been discussed previously \cite{Rutgers_PhysFluids_1996,Rutgers_RevSciInstr_2001}.
Briefly, our soap-film channel is made of two parallel nylon wires,
which are connected to two soap solution reservoirs, one at the top
and the other at the bottom of the channel. To create a soap film,
we first let the soap solution flow along the two wires, and then
the wires are separated from each other to form a film. The channel
width in the current measurement is set constant at $5\,\text{cm}$.
The soap solution collected by the lower reservoir is pumped back
to the top reservoir, resulting in a long-lasting film once it is
initiated. The film flow speed ($1.5<V_{F}<2.5\,\text{m/s}$) and
the thickness ($0<h\lesssim10\,\text{\ensuremath{\mu}m}$) can be
varied by regulating the solution flux by a valve at the injection
point at the top of the channel.

The soap solution used consists of $2\%$ (in volume) liquid detergent
(Dawn) and 98\% distilled water. The kinematic viscosity of the soap
solution, measured by a glass capillary viscometer (Cannon Instrument
Co.), is $\nu\simeq0.012\,\text{c\ensuremath{m^{2}}/s}$, close to
water. The surface tension of the liquid/air interface of the bulk
solution was determined by a Du No\"uy tensiometer (CSC Scientific)
as $\sigma\simeq32\pm1\,\text{erg/cm}^{2}$. Measurements of $\sigma$
as a function of the concentration of the liquid detergent indicate
that the critical micellar concentration of our soap solution is well
below $0.1\%$. As will be discussed below, a wave speed measurement
in the soap film indicates that the surface tension of the film is
about the same as measured on the free surface of the bulk liquid,
indicating that at $2\%$ concentration, the surfaces of the film
are fully covered by the surfactant. 

Micron-sized water droplets are created by using an inkjet printer
cartridge (NEC model \#30-060). This cartridge works by electrically
heating water near a nozzle, which consists of a small heater with
a $30\,\Omega$ internal resistance. When a pulse of electric current
is applied, the rapid expansion of water near the heater causes a
droplet to be ejected with a high speed. By using a custom-made computer
program and a current driving circuit, a rectangular electric pulse
of $7.5\,\text{V}$ is applied to the heater for $\sim10\,\text{\ensuremath{\mu}s}$.
The pulse is repeated periodically and is controlled by the computer.
In a typical measurement $\sim10$ ballistic droplets are created
per second and their motion is followed by video imaging.

We determined the droplet radius by weighing the cartridge before
and after ejecting $9\times10^{4}$ droplets. The decrease in the
cartridge mass was $6.5\pm1$ mg, averaged over several trials. This
yields the average mass per droplet $72\pm11$ ng or a radius $R=26\pm1$
$\text{\ensuremath{\mu}m}$. Independently, using fast video imaging,
the\textcolor{black}{{} terminal} velocity $v_{t}=7.6\pm1.1\,\text{cm/s}$
of the droplet was determined. For a small Reynolds number in air,
\textcolor{black}{$\text{R\ensuremath{e_{a}}}(\equiv2Rv_{t}/\nu_{a})<1$,
}$v_{t}$ of a sphere is approximately given by $v_{t}\simeq\frac{mg}{6\pi\eta_{a}R}$,
where $\eta_{a}$ and $\nu_{a}$ are respectively the shear and the
kinematic viscosity of air. This yields a droplet radius of $25.4\pm1.7\,\text{\ensuremath{\mu}m}$,
which is in good agreement with the weighing method. In what follows,
we will use $R_{0}=26\,\text{\ensuremath{\mu}m}$ as the radius of
our droplets.

To study the interactions with a soap film, a stream of droplets generated
by the inkjet nozzle is aimed normal to the film. For convenience
of a latter discussion, a coordinate system is set up such that the
initial velocity is along the $x$-axis and the film flows in $\hat{y}$
direction as depicted in Fig. \ref{cap:expsetup}. The impact velocity
$v_{i}$ is varied by adjusting the distance between the nozzle and
the film. After the collision, the droplet either tunnels through
with a non-zero $v_{x}$ or is absorbed by the film with $v_{x}=0$.
In this experiment, we measured the velocity components $(v_{x},v_{y})$
after tunneling, as a function of the film thickness $h$ and $v_{i}$.
To visualize the droplet's trajectory, a high-intensity (300 W) halogen
lamp and a high-speed video camera (Phantom V5, Vision Research) were
used. The camera operates at several thousand frames per second, allowing
droplet velocity to be measured reliably.

\begin{figure}
\includegraphics[width=8.5cm]{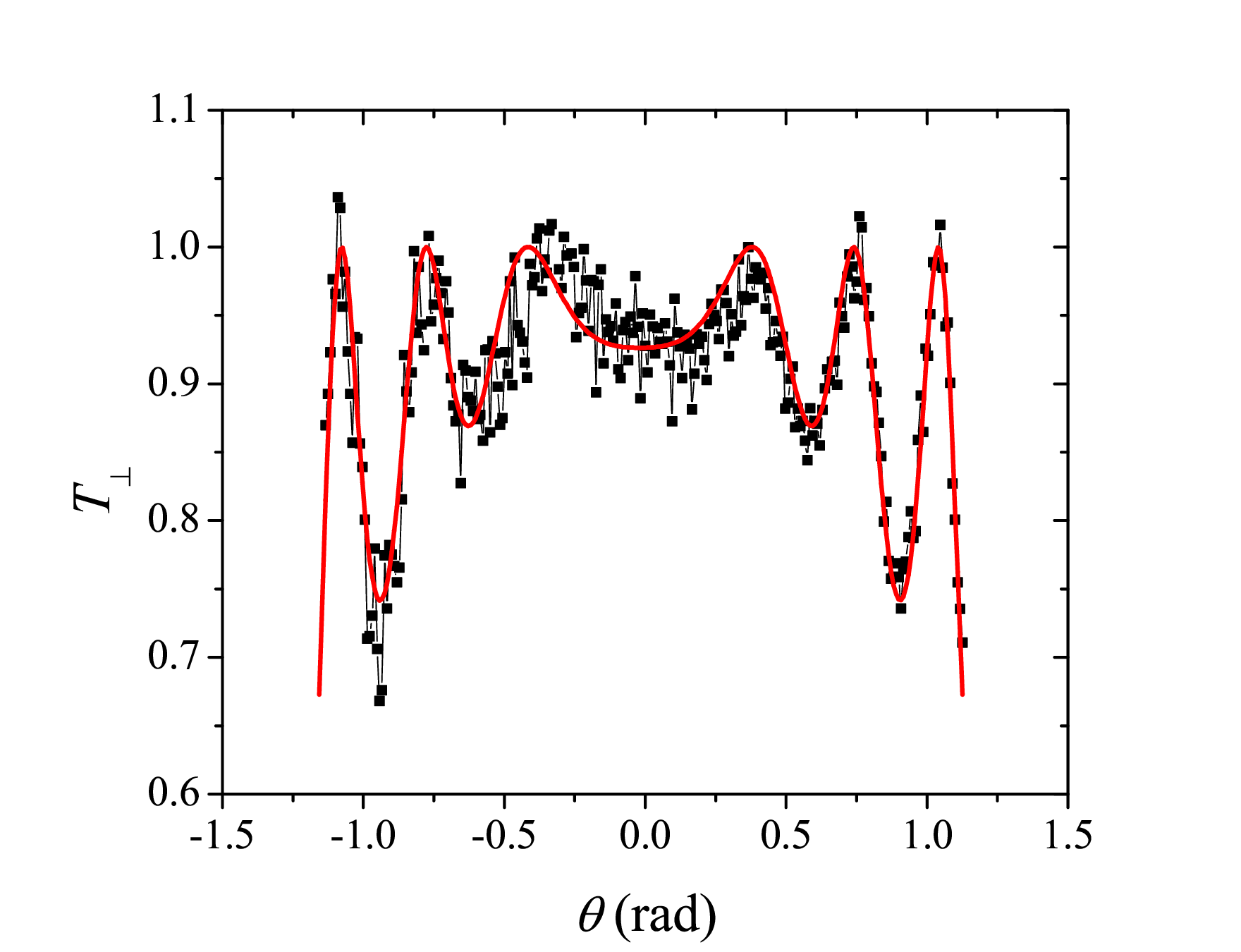}

\caption{A typical optical transmission measurement in a flowing film. The
data (dots) show transmittance $T_{\perp}$ as a function of the incident
angle $\theta$. We found that a soap film can be well approximated
by a dielectric slab with a constant thickness $h$ and a refractive
index $n=1.33$. This allows $T_{\perp}$ to be calculated rigorously.
The solid line is a fit to the slab model of Eq. \eqref{eq:Transmittance},
resulting in $h\simeq2.56\,\text{\ensuremath{\mu}m}$ in this case.
\label{cap:slabmodel}}

\end{figure}

To determine the film thickness $h$, we measured the optical transmittance
of the $p$-polarized light $T_{\perp}$ at $\lambda=685\,\text{nm}$
as a function of incident angle $\theta$ of a semiconductor laser.
A rotation stage was built to allow the laser and a photodiode to
be rotated synchronously on two separate arms. A computer controlled
stepping motor drives the rotation stage, allowing a wide range of
incident angle to be scanned rapidly, i.e., $-70\textdegree\le\theta\le+70\textdegree$
in $\sim5\,\text{s}$. We modeled the soap film as a dielectric slab
of thickness $h$ with a refractive index $n=1.33$. A calculation
shows that $T_{\perp}$ is given by \cite{Hecht} \begin{equation}
T_{\perp}=\left[1+\left(\frac{2r_{\perp}}{1-r_{\perp}^{2}}\right)^{2}\sin^{2}\left(\delta/2\right)\right]^{-1},\label{eq:Transmittance}\end{equation}
where $r_{\perp}=\frac{\cos\theta-\sqrt{n^{2}-\sin^{2}\theta}}{\cos\theta+\sqrt{n^{2}-\sin^{2}\theta}}$
and $\delta=\frac{4\pi h}{\lambda}\sqrt{n^{2}-\sin^{2}\theta}$. As
shown in Fig. \ref{cap:slabmodel}, this slab model works quite well
for our soap films, and $h$ can be accurately determined.

\section{Results and Discussions}

\begin{figure}
\begin{centering}
\includegraphics[width=8.5cm]{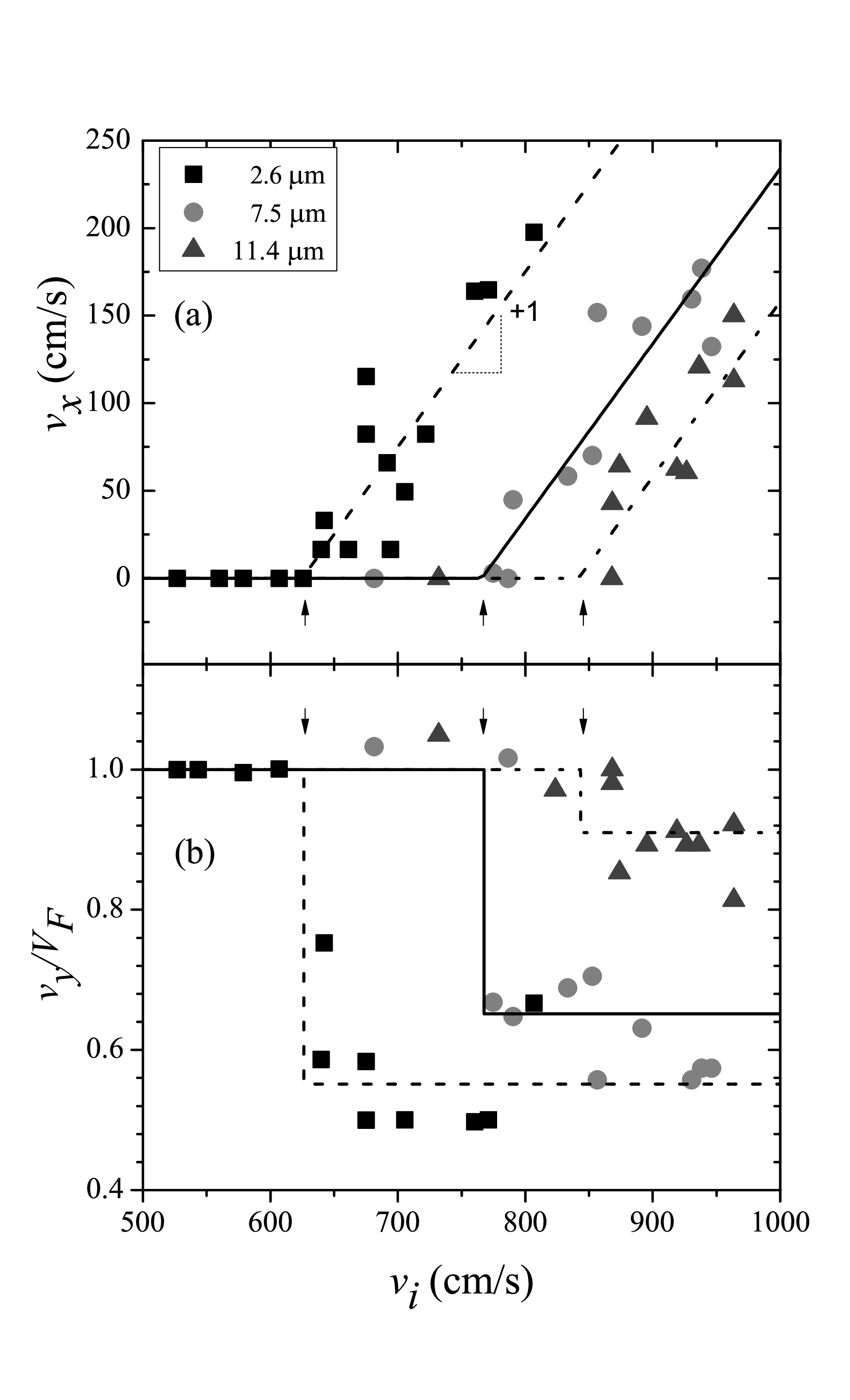}
\par\end{centering}

\caption{The droplet velocity $(v_{x},v_{y})$ after impact as a function of
the impact velocity $v_{i}$. The measurements were carried out for
different film thicknesses: $h=$2.6 $\text{\ensuremath{\mu}m}$ (squares),
7.5 $\text{\ensuremath{\mu}m}$ (\textcolor{black}{circles), and 11.4
$\text{\ensuremath{\mu}m}$ (triangles}). In (a), the $x$-component
of the velocity $v_{x}$ is plotted against $v_{i}$. It is shown
that there exists a critical velocity $v_{C}$ for each film thickness
as indicated by the vertical arrows. In (b), the normalized $y$-component
of the velocity $v_{y}/V_{F}$ is plotted against $v_{i}$. Here we
found that $v_{y}/V_{F}$ has two discrete values for a given $h$.
It is either +1 when the droplet merges with the film, or $\epsilon\le1$
when the droplet tunnels through the film. \label{cap:vsimpactvelocities}}

\end{figure}

\begin{figure}
\includegraphics[width=8.5cm]{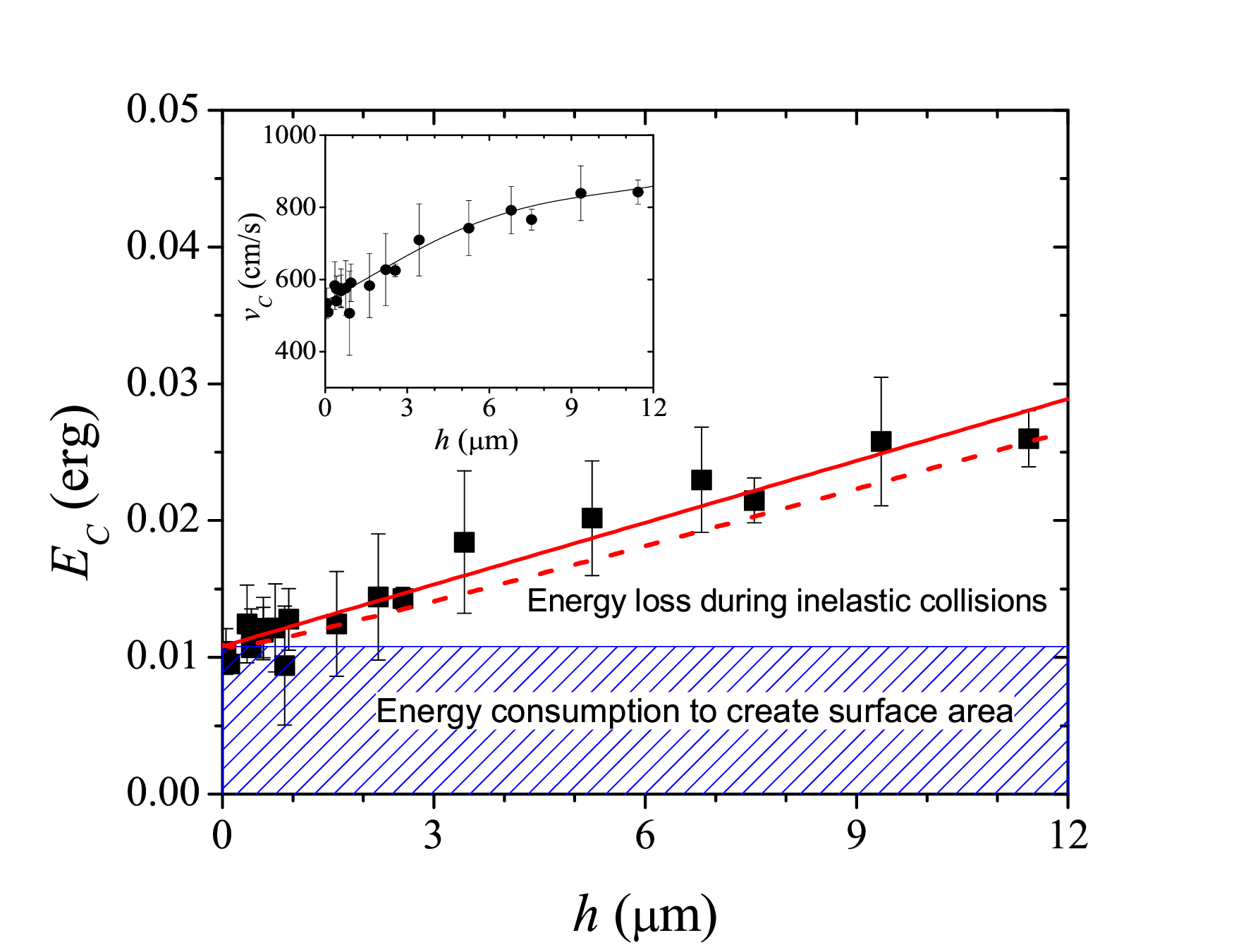}

\caption{\textcolor{black}{The critical velocity $v_{C}$ and the energy barrier
$E_{C}\equiv\frac{1}{2}mv_{C}^{2}$ vs. the film thickness $h$.}
In the inset, the critical velocity $v_{C}$ is plotted against $h$.
The solid line is a guide to the eyes. We define $E_{C}=\frac{1}{2}mv_{C}^{2}$,
which is plotted as solid squares in the main graph. It is shown that
$E_{C}(h)$ is approximately linear in $h$, but it does not vanish
as $h\rightarrow0$, indicating that a finite film deformation energy
is required for tunneling. The solid line is a calculation based on
Eq. \eqref{eq:Ecvsh}, and the dotted line is a calculation based
on Eq. \eqref{eq:Ecvshoriginal}. (see text for more details). The
data points with $h<2\,\text{\ensuremath{\mu}m}$ were measured using
static soap films ($V_{F}=0$). \label{cap:h-dependence-w-theory-x}}

\end{figure}

Figure \ref{cap:vsimpactvelocities}(a) displays the horizontal velocity
component of the droplets after the impact on a soap film $v_{x}$
as a function of $v_{i}$. The measurements were repeated for films
of different thicknesses $h$. It is shown that there exists a critical
velocity $v_{C}$; for $v_{i}<v_{C}$, the droplet loses its horizontal
momentum ($v_{x}=0$) and it flows with the soap film after impact.
On the other hand, for $v_{i}>v_{C}$, the droplet is able to tunnel
through the film with $v_{x}>0$. Although the data near the threshold
is somewhat noisy, the critical velocity $v_{C}$ can be determined
without much ambiguity by extrapolating the data below and above the
threshold. As delineated in the inset of Fig. \ref{cap:h-dependence-w-theory-x},
our measurements indicate that $v_{C}$ increases with $h$, suggesting
that the energy barrier for tunneling becomes greater for a larger
$h$. We also found that $v_{C}$ does not vanish when $h\rightarrow0$
but approaches a finite value $v_{C0}\sim520\,\text{cm/s}$, which
translates to a minimum energy $E_{min}(\equiv\frac{1}{2}mv_{C0}^{2})\simeq0.01\,\text{erg}$.
This suggests that a considerable fraction of the tunneling energy
$E_{C}$ is in the deformation of the soap film. Another conspicuous
feature seen in Fig. \ref{cap:vsimpactvelocities}(a) is that after
tunneling, $v_{x}$ is linearly proportional to the impact velocity
$v_{i}$, and the proportionality constant is to a good approximation
unity for all different $h$.

\begin{figure}
\includegraphics[width=8.5cm]{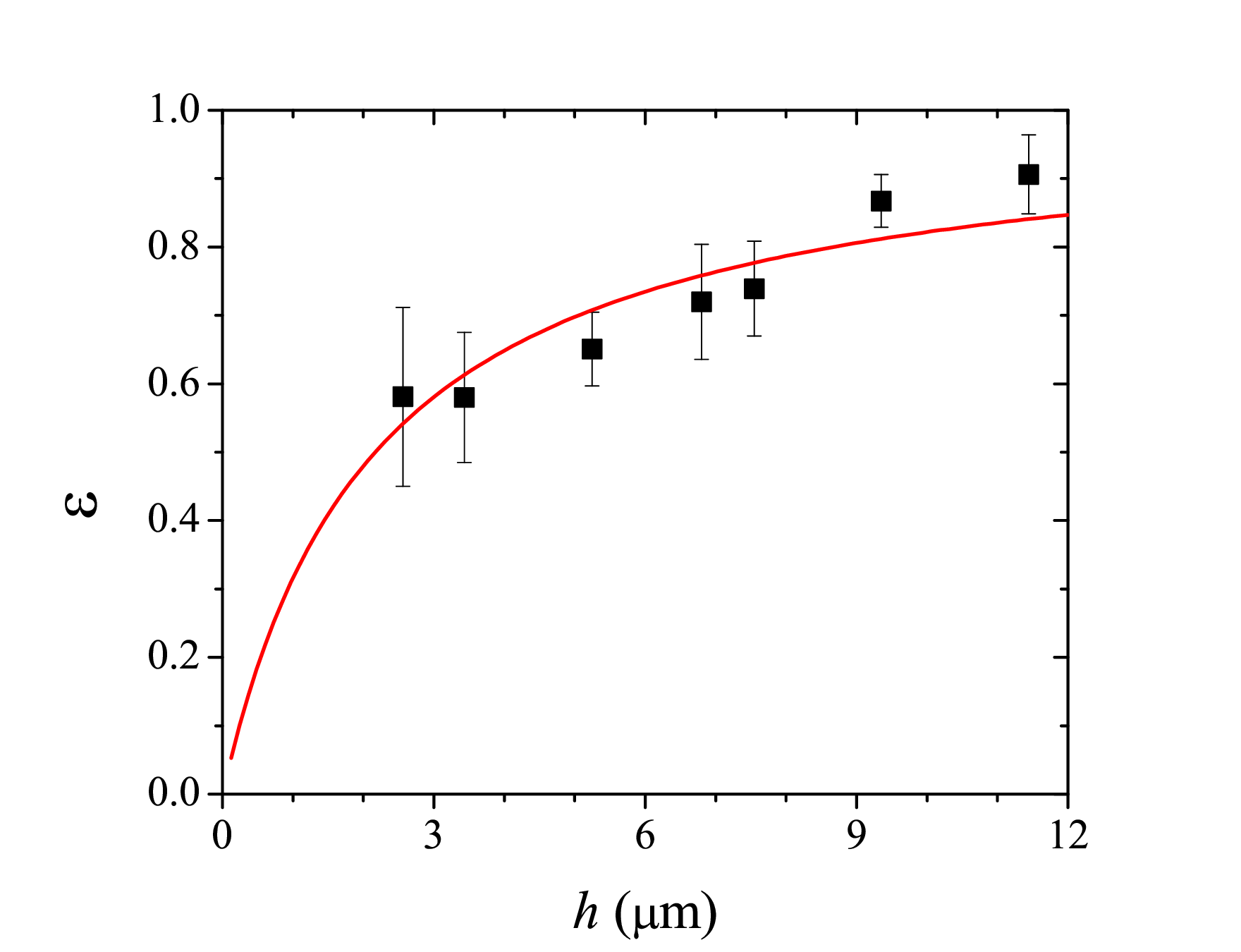}

\caption{The normalized vertical velocity after tunneling $\epsilon=v_{y}/V_{F}$
vs. the film thickness $h$. It is observed that as $h$ increases,
$\epsilon$ becomes larger, indicating that the droplet gains a linear
momentum in the $\hat{y}$ direction. Asymptotically, $v_{y}\rightarrow V_{F}$
(or $\epsilon\rightarrow1$) as $h\rightarrow\infty$ as expected.
The solid line is a calculation based on Eq. \eqref{eq:vyvsh-2} derived
from an inelastic collision model. \label{cap:h-dependence-w-theory-y}}

\end{figure}

\textcolor{black}{In addition to the linear momentum exchange between
the droplet and the soap film in the horizontal direction, the momentum
exchange in the vertical }direction\textcolor{black}{{} is equally significant.
Fig. \ref{cap:vsimpactvelocities}(b) displays normalized $y$-component
velocity $v_{y}/V_{F}$ as a function of the impact velocity $v_{i}$.
It is shown that if $v_{i}<v_{C}$, $v_{y}/V_{F}=1$, i.e. the droplet
moves with the film. However, for $v_{i}>v_{C}$, the tunneling causes
the droplet to gain a momentum in the $\hat{y}$ direction, but in
general $v_{y}<V_{F}$. Parameterizing the change in the $y$-component
velocity of the droplet before and after tunneling $\epsilon=v_{y}/V_{F}$,
we found that $\epsilon$ is nearly independent of $v_{i}$ but it
depends on the film thickness $h$ as delineated in Fig. \ref{cap:h-dependence-w-theory-y}.
It shows that when the droplet passes through a thicker film, it gains
more momentum in the $\hat{y}$ direction than passing through a thin
one. Asymptotically, one expects that $\epsilon\rightarrow1$ as $h\rightarrow\infty$,
which is consistent with our observation. We found that a tunneling
droplet readily picks up the $y$-momentum from the film. For instance,
at the current experimental condition, $v_{y}$ is $\sim60\%$ of
$V_{F}$ for a film as thin as a few microns and is $\sim90\%$ for
$h\simeq10\,\text{\ensuremath{\mu}m}$. }

\begin{figure}
\includegraphics[width=8.2cm]{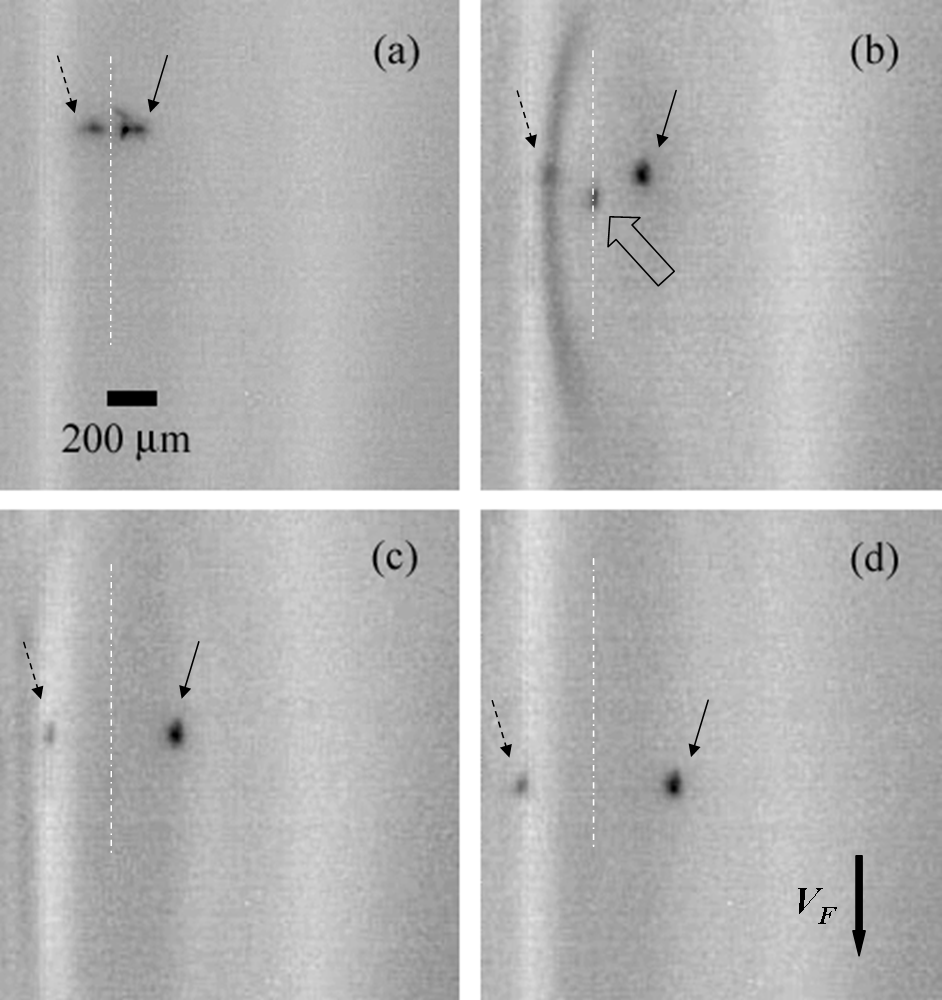}

\caption{\textcolor{black}{The tunneling dynamics of a water droplet. The sequence
of video images (a-d) were taken at an equal time interval with $t=0$,
$0.26$, $0.52$, and $0.78$ ms. Here the penetrating droplet and
its mirror image are clearly visible and are marked by the pair of
arrows, where the dotted lines depict the plane of reflection.  As
a function of time, the droplet/image pair moves together in the vertical
direction, but they moves apart from each other in the horizontal
direction. This allows a precise determination of $v_{x}$ and $v_{y}$.
Shortly after the impact, an elongated dark region can be identified
in (a), showing that the film is stretched by the droplet. A moment
later (b), a scar, which is indicated by a large arrow, is left behind
in the film. The scar moves with $V_{F}$, which is faster than the
droplet velocity $v_{y}$. This yields a precise determination of
the velocity difference $v_{y}-V_{F}$ or $\epsilon$. Interestingly,
the scar in (b) disappears rapidly and is no longer observable in
(c). Also seen in (b) is the surface wave (the dark band) that propagates
radially outwards from the impact point.\label{cap:passingdrop}}}

\end{figure}

\begin{figure}
\includegraphics[width=8.2cm]{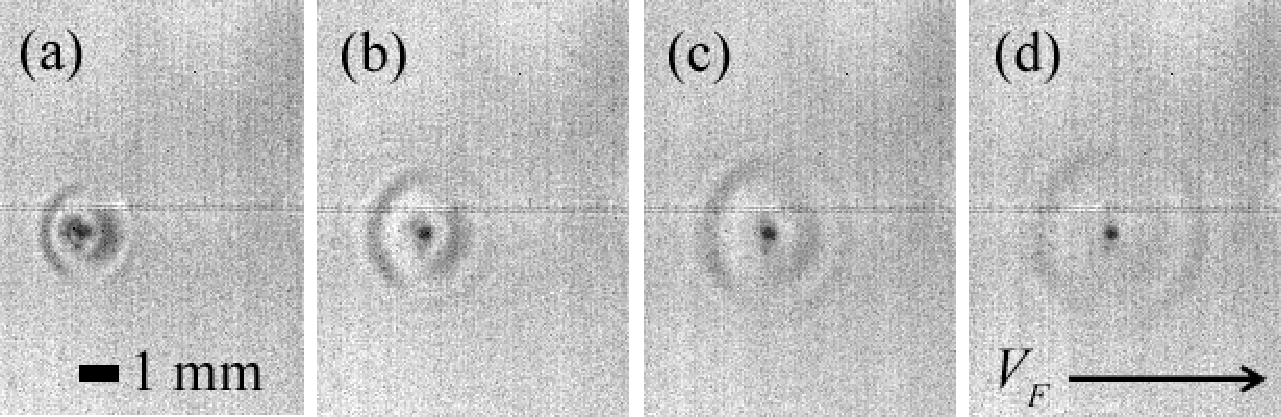}

\includegraphics[width=8.5cm]{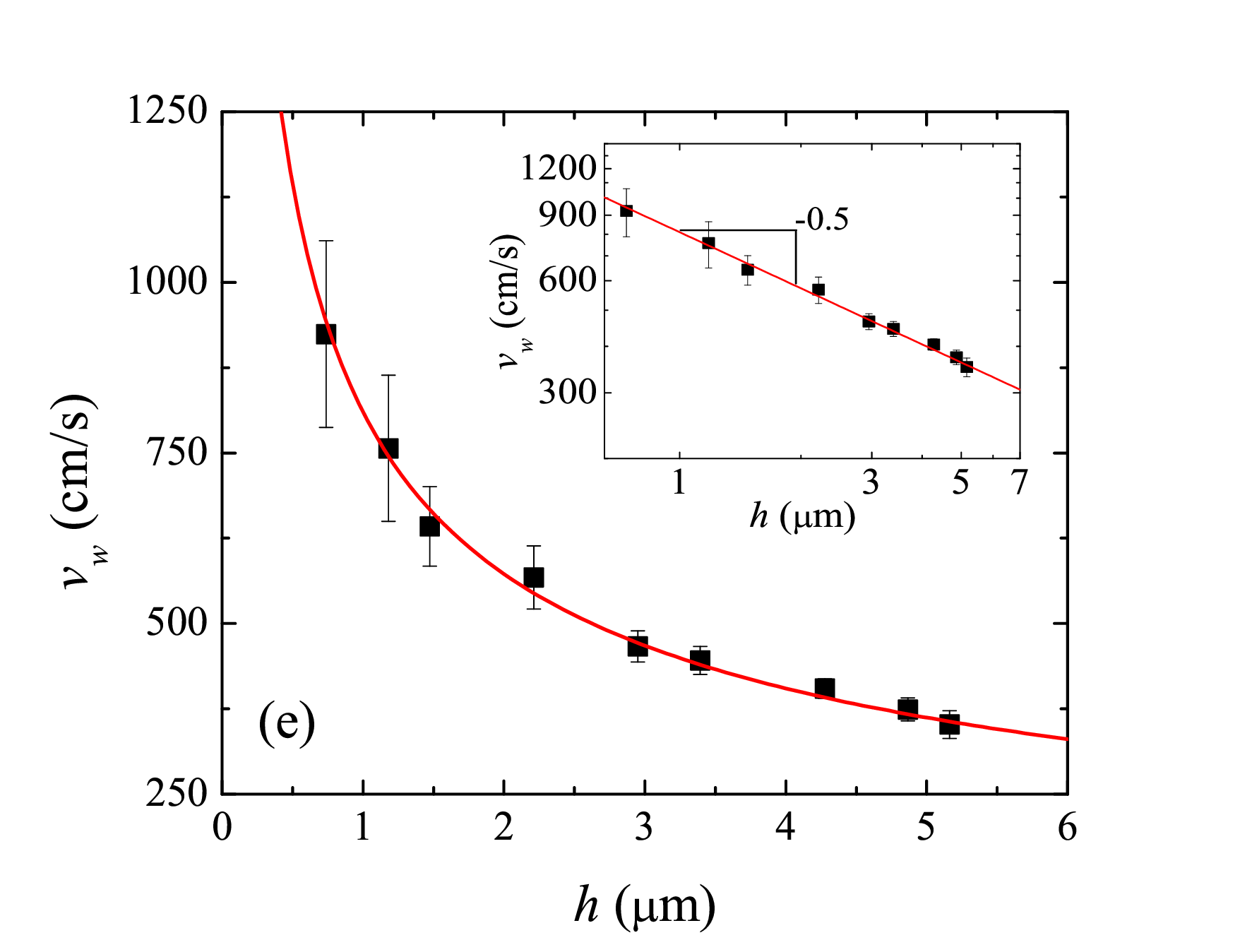}

\caption{A wave generated by an impacting droplet. Images (a-d) were taken
sequentially at an equal interval, corresponding to $t=0$, $0.2$,
$0.4$, and $0.6$ ms, respectively. The impact velocity of the droplet
is less than $v_{C}$ so that it merges with the soap film. The droplet
appears as a dark spot at the center of the expanding circular wavefront.
In (e), the wave speed $v_{w}$ in the soap films is measured as a
function of $h$. \textcolor{black}{The scaling relation $v_{w}\propto h^{-1/2}$
is delineated in the inset, where }the solid line is a fit to $v_{w}=\sqrt{\frac{2\sigma}{\rho_{w}h}}$.\label{cap:wavespeed}}

\end{figure}

\textcolor{black}{To understand the physical origin of the energy
barrier in the tunneling process,} we examined carefully the impact
dynamics using fast video imaging while the film was illuminated by
a monochromatic sodium lamp. Fig. \ref{cap:passingdrop} (a-d) displays
four consecutive images of a droplet shortly after it had impacted
on a soap film of thickness $h=4\,\text{\ensuremath{\mu}m}$. The
time interval between the images was fixed at $0.26\,\text{ms}$.
Here, the droplet as well as its mirror image on the soap film are
clearly visible. Also visible in (b) is a scar created by the droplet,
but the scar disappeared in (c), indicating that its lifetime is less
than $0.26\,\text{ms}$.\textcolor{red}{{} }\textcolor{black}{Because
of the short length and time scales involved, we believe that the
scar region is associated with strong vorticity, which dissipates
energy.} The use of the monochromatic light also allowed us to observe
waves created by the impact. Here, the wavefront appears as a dark
band in (b) that propagates radially outwa\textcolor{black}{rd. In
Fig. \ref{cap:wavespeed}(a-d), another sequence of images is captured
at an equal interval $0.2\,\text{ms}$ by setting the camera normal
to the surface of the soap film. Here the droplet is merged with the
film because it impacts the film at the velocity lower than $v_{C}$.
The merging droplet is visible as a dark spot at the center of the
circular wavefront and is carried downstream by the film. As it moves
downstream, the wavefront propagates radially outward at a constant
velocity $v_{w}$. By investigating images such as ones in Fig. \ref{cap:wavespeed},
the wa}ve speed $v_{w}$ on the soap film can be determined as a function
of $h$, which is displayed as solid squares in Fig. \ref{cap:wavespeed}(e).
The figure shows that $v_{w}$ increases rapidly as $h$ decreases;
for a small thickness, $h<1\,\text{\ensuremath{\mu}m}$, $v_{w}$
can be as large as $10\,\text{m/s}$. A liquid film in general can
support two different types of waves, known as the symmetric and the
anti-symmetric waves \cite{Taylor1959b}. For the symmetric wave,
the two surfaces of the film move out of phase with respect to each
other, and is also called peristaltic mode of oscillations (see Fig.
\ref{cap:diagram}(b)). For the anti-symmetric wave, both surfaces
undulate in phase (see Fig. \ref{cap:diagram}(c)). In the absence
of surfactants, the restoring force of both waves are due to the surface
tension $\sigma$ but the dispersion relations are different for the
two cases because of different mass distributions in the film. The
anti-symmetric wave is non-dispersive with a constant velocity given
by $v_{a}=\sqrt{\frac{2\sigma}{\rho_{w}h}}$, and the symmetric wave,
on the other hand, is dispersive with a velocity $v_{s}=k\sqrt{\frac{\sigma h}{2\rho_{w}}}$
that depends on a wavenumber $k$. Since $v_{s}/v_{a}=kh/2$, it is
expected that $v_{a}\gg v_{s}$ in the long-wavelength limit. In the
presence of surfactants, the restoring force for the peristaltic mode
is dominated by the surface (or Marangoni) elasticity $E\equiv A\frac{d\sigma}{dA}$,
where $A$ is a surface area, instead of $\sigma$. It is shown by
Lucassen \cite{Lucassen_1970} that in the long-wavelength limit,
the elastic wave is also non-dispersive with a propagation speed $v_{e}=\sqrt{\frac{2E}{\rho_{w}h}}$.
In our soap film therefore there is a degeneracy in that both the
anti-symmetric capillary wave and the symmetric elastic wave are possible,
and both scale with the film thickness as $h^{-1/2}$. We found that
our experimental data in Fig. \ref{cap:wavespeed} can be well described
by the mathematical form $v_{w}=\sqrt{\frac{2c}{\rho_{w}h}}$ (see
the inset), where $c$ is an adjustable parameter. Using $\rho_{w}=1\,\text{g/cm}^{3}$,
a fitting procedure yields $c\simeq32.7\pm2.6\,\text{erg/cm}^{2}$,
which matches very well with the surface tension measurement ($\sigma=32\,\text{erg/cm}^{2}$)
using the Du No\"uy ring method. Thus, unless it is an amazing coincidence,
where Marangoni elasticity $E$ is nearly identical to $\sigma$ for
our soap film, we believe that the dominant wave mode in the film
created by the droplet impact is the anti-symmetric wave. 

At the impact point (see Fig. \ref{cap:passingdrop}(a)), it is observed
that the droplet deforms the soap film locally, forming a cylindrical
pouch \textcolor{black}{a few droplet diameters long}. Because of
smallness of the droplet and fast dynamics, it was not feasible to
follow the spatiotemporal evolution of the cylindrical pouch. However,
inspections of a large number of video images reveal that the longest
pouch is $\sim4R_{0}$, indicating that in order to tunnel through,
the soap film must be stretched into a long cylinder with an excess
surface area of $\sim14\pi R_{0}^{2}$, where $R_{0}$ is the radius
of the water droplets. A naive calculation using $\sigma=32\,\text{erg/cm}^{2}$
indicates that this corresponds to a surface energy of $9.5\times10^{-3}\,\text{erg}$.
Considering the Marangoni effect, we expect that the energy requirement
may be greater.

\textcolor{black}{A question that arises naturally is what determines
the maximum length of the cylinder.} A related phenomenon is the Plateau-Rayleigh
instability where a uniform circular jet of fluid breaks up into a
stream of droplets \cite{Chandrasekhar}. For the anti-symmetric undulation
to be the dominant mode of oscillations in our soap film, the instability
of a cylindrical soap film is similar to the Plateau-Rayleigh problem
with the simple modification of replacing the surface tension $\sigma$
by $2\sigma$, due to the presence of two liquid-air interfaces of
the film. It follows that the fastest growing wavenumber $k_{max}$
of the axial undulation is given by $k_{max}\simeq0.7/R_{0}$, corresponding
to $\lambda_{max}(\equiv2\pi/k_{max})\simeq9R_{0}$ \cite{Chandrasekhar}.
We note that our experimentally observed pouch length $4R_{0}$ is
about a half of $\lambda_{max}$, which makes physical sense since
the front of the pouch, where the water droplet locates, must be the
anti-node and the location of the pinch-off must be the node given
by $(n+\frac{1}{2})\lambda_{max}$ where $n=0,1,2,\ldots$. Our observation
corresponds to $n=0$ (see Fig. \ref{cap:diagram}(a)). This is possible
if the pinch-off time $\tau$ is shorter than the stretching time
$(\sim\lambda_{max}/2U)$, because otherwise a longer pouch will be
produced and it will break in multiple locations, which was not observed.
Quantitatively, this scenario also makes sense since according to
our measurement the stretching time is $\lambda_{max}/2U\simeq12\,\text{\ensuremath{\mu}s}$
and the pinch-off time $\tau$ can be estimated by the growth rate
of the Plateau-Rayleigh instability $\tau\equiv\sqrt{\rho_{a}R_{0}^{3}/2\sigma}\simeq0.52\,\text{\ensuremath{\mu}s}$.
Thus the condition $\tau\ll\lambda_{max}/2U$ is satisfied. The emerging
physical picture is that during transmission, a piece of soap film
is extruded by the fast moving droplet. At the same time an axial
undulation grows rapidly on the stretched cylindrical film, and the
cylinder closes off at its base once $\lambda_{max}/2$ is reached.
As we will show below, in order to explain the kinematics of the tunneling
droplet, a small mass must be transferred from the film to the droplet
and the size of such a mass can be determined from our measurements.

\section{Calculations}

It would be desirable to compare our experimental observations with
theoretical predictions. Unfortunately such theory is not currently
available. A back-of-the-envelope calculation shows that our measurements
were carried out in a hydrodynamic regime where Reynolds number $\text{Re}=\frac{2\rho_{w}R_{0}v_{i}}{\eta_{w}}\gtrsim260$
and the Weber num\textcolor{black}{ber $\text{We}=\frac{2\rho_{w}R_{0}v_{i}^{2}}{\sigma}\gtrsim44$
are} both large. Thus the kinetic energy or the inertia effect overwhelms
energy dissipation and the capillary effect. In the following we propose
a heuristic model that can account for some key features of our observations. 

\begin{figure}
\includegraphics[width=8.5cm]{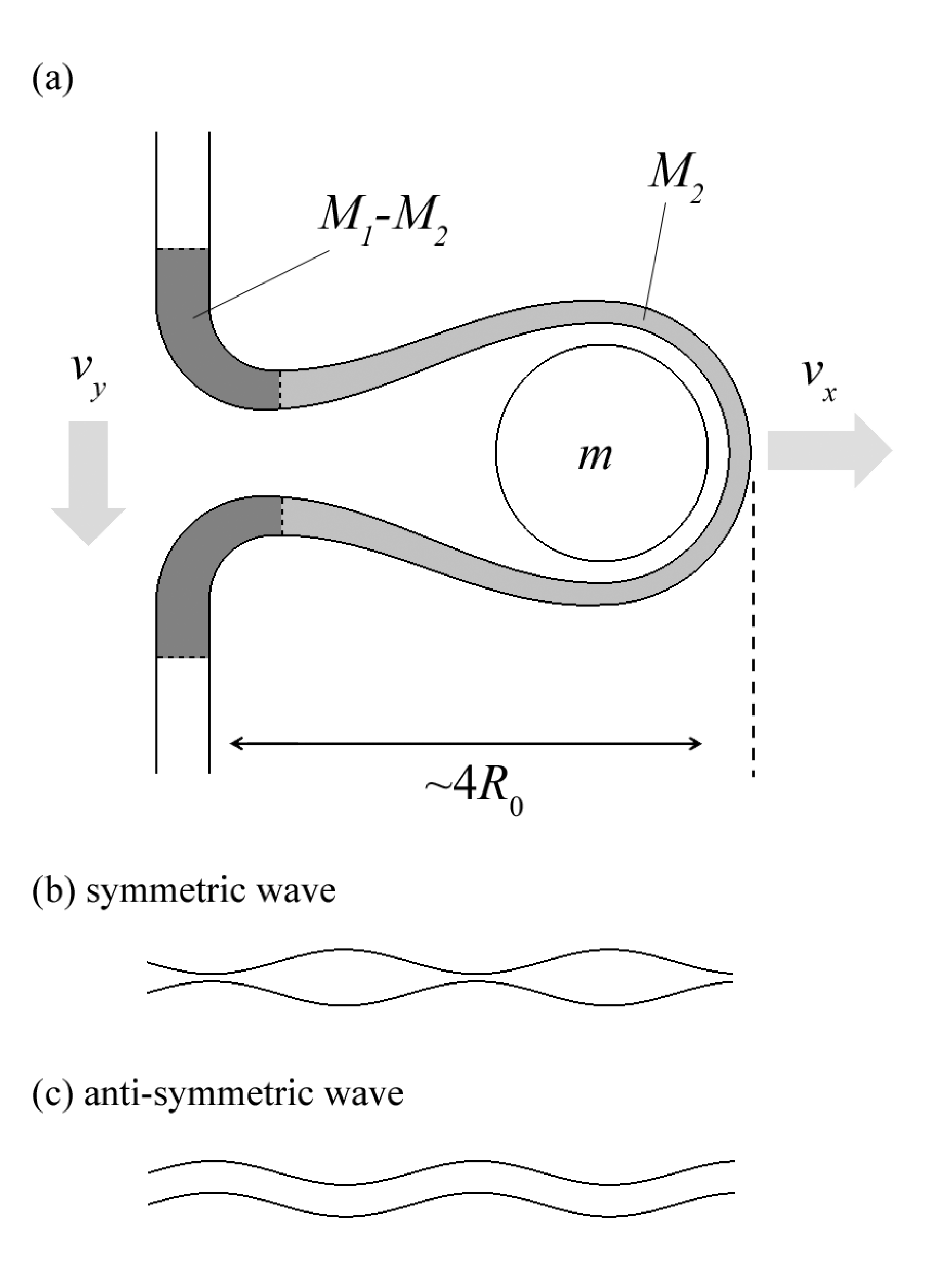}

\caption{The schematics depicting different conformations of a soap film. In
(a), the film is stretched by a ballistic droplet, where $M_{2}$
will eventually engulf droplet $m$, and $M_{1}$ provides a $y-$momentum
to the droplet. (b) and (c) are two possible wave modes in a soap
film.\label{cap:diagram}}

\end{figure}

It is apparent that when a collision takes place, only a fraction
of film mass in the neighborhood of the impact is involved in the
interaction. Thus an effective mass $M_{1}$ and its corresponding
size $R_{1}=\sqrt{M_{1}/\pi\rho_{w}h}$ may be specified. After the
collision, a part of $M_{1}$, called $M_{2}$, is transferred to
the water droplet, increasing its mass from $m$ to $m+M_{2}$. The
emerging droplet travels with velocity $(v_{x},v_{y})$ and the remaining
mass $M_{1}-M_{2}$ travels along the film with velocity $(v_{x}',v_{y}')$
(see Fig. \ref{cap:diagram}(a)). The presence of $M_{2}$ allows
a second length scale $R_{2}=\sqrt{M_{2}/\pi\rho_{w}h}$ to be specified.
We treat our problem as an inelastic collision in which linear momentum
is conserved but not energy. 

The linear momentum conservation demands \begin{equation}
mv_{i}=(m+M_{2})v_{x}+(M_{1}-M_{2})v_{x}',\end{equation}
 \begin{equation}
M_{1}V_{F}=(m+M_{2})v_{y}+(M_{1}-M_{2})v_{y}'.\end{equation}
\textcolor{black}{A}t the moment of separation, when the dressed droplet
($m+M_{2}$) becomes detached from the rest of the film, we expect
that $v_{x}>0$ but $v_{x}'\simeq0$. For the inelastic collision,
one also expects $v_{y}=v_{y}'$ at the separation point. Solving
the above equations, we find \begin{equation}
v_{x}=\frac{m}{m+M_{2}}v_{i},\label{eq:v_x}\end{equation}
\begin{equation}
v_{y}=v_{y}'=\frac{M_{1}}{m+M_{1}}V_{F}.\label{eq:v_y}\end{equation}
The total kinetic energy $KE_{f}$ of the droplet and the film after
collision is given by\begin{eqnarray}
KE_{f} & = & \frac{1}{2}(m+M_{2})(v_{x}^{2}+v_{y}^{2})+\frac{1}{2}(M_{1}-M_{2})(v_{y}'^{2}),\nonumber \\
 & = & \frac{1}{2}\frac{m}{m+M_{2}}mv_{i}^{2}+\frac{1}{2}\frac{M_{1}}{m+M_{1}}M_{1}V_{F}^{2}.\end{eqnarray}
Neglecting the deformation of the film, this is all the mechanical
energy of the system (droplet plus the soap film) that remains after
the collision. It is clear that the collision is inelastic since

\begin{eqnarray}
\Delta KE( & \equiv & KE_{i}-KE_{f})\nonumber \\
 & = & \frac{1}{2}m\left(\frac{M_{2}}{m+M_{2}}v_{i}^{2}+\frac{M_{1}}{m+M_{1}}V_{F}^{2}\right)\geq0,\end{eqnarray}
where $KE_{i}=\frac{1}{2}mv_{i}^{2}+\frac{1}{2}M_{1}V_{F}^{2}$ is
the total kinetic energy of the system before the collision. Physically,
$\Delta KE$ is the amount of energy ultimately dissipated by the
creation of vorticity in the fluid. Since both $M_{1}$ and $M_{2}$
are proportional to $h$, we found that energy dissipation vanished
when $h\rightarrow0$. In other words, no vorticity can be created
in a very thin film so the physics of tunneling becomes a purely potential
flow problem. For a successful transmission, the energy consideration
therefore requires that the initial energy of the droplet should be
greater than the sum of the energy dissipation $\Delta KE$ and the
film deformation $E_{min}$ with the result: $KE_{i}\ge\Delta KE+E_{min}$
(or $KE_{f}\ge E_{min}$). This yields

\begin{equation}
\frac{1}{2}mv_{i}^{2}\ge E_{min}\left(1+\frac{M_{2}}{m}\right)-\frac{1}{2}mV_{F}^{2}\left(\frac{m+M_{2}}{m+M_{1}}\right),\end{equation}
or the critical energy of the droplet $E_{C}=\frac{1}{2}mv_{C}^{2}$
as: \begin{equation}
E_{C}=E_{min}\left(1+\frac{M_{2}}{m}\right)-\frac{1}{2}mV_{F}^{2}\left(\frac{m+M_{2}}{m+M_{1}}\right).\label{eq:Ecvshoriginal}\end{equation}
This equation correctly predicts that the motion of the film ($V_{F}\neq0$)
lowers the energy barrier of tunneling. For a small $V_{F}$, e.g.,
in our experiment $(V_{F}/v_{C})^{2}<\text{0.15}$, one may neglect
the last term to obtain, \begin{equation}
E_{C}\simeq E_{min}\left(1+\frac{M_{2}}{m}\right)=E_{min}\left(1+\frac{3}{4}\alpha_{2}^{2}\frac{h}{R_{0}}\right),\label{eq:Ecvsh}\end{equation}
where $\alpha_{2}=R_{2}/R_{0}$ is a constant. Eq. \eqref{eq:Ecvsh}
is consistent with our observation in that $E_{C}$ is linear in $h$
with a finite intercept (see Fig. \ref{cap:h-dependence-w-theory-x}).
Using the known parameters of our soap film ($E_{min}\simeq0.01\,\text{erg}$
and $R_{0}=26\,\text{\ensuremath{\mu}m}$), we found $\alpha_{2}\simeq2.2\pm0.1$.
Interestingly, this value of $\alpha_{2}$ implies a rather uniform
coating of the penetrating droplet by the film of thickness $h$,
i.e., $\Delta V\simeq4\pi R_{0}^{2}h$. In Fig. \ref{cap:h-dependence-w-theory-x}
we showed that Eq. \eqref{eq:Ecvsh} (solid line) and Eq. \eqref{eq:Ecvshoriginal}
(dotted line) both agree reasonably with our measurements.

According to Eq. \eqref{eq:v_y}, the emerging droplet will have a
velocity in vertical direction

\begin{equation}
v_{y}=\frac{M_{1}}{m+M_{1}}V_{F}=\frac{V_{F}}{1+\frac{4}{3}\left(\frac{R_{0}}{\alpha_{1}^{2}h}\right)},\label{eq:vyvsh-2}\end{equation}
where $\alpha_{1}=R_{1}/R_{0}$. This equation yields the correct
asymptotic behavior, $\epsilon(h)(\equiv v_{y}/V_{F})\rightarrow1$
as $h\rightarrow\infty$, as seen in Fig. \ref{cap:h-dependence-w-theory-y}.
Using $\alpha_{1}$ as an adjustable parameter, a fitting procedure
gives $\alpha_{1}\simeq4.0\pm0.4$, which is delineated by the solid
line in Fig. \ref{cap:h-dependence-w-theory-y}.

\section{Conclusion }

Using well-controlled micron-sized ballistic droplets generated by
an inkjet cartridge, we have characterized the energy requirement
for tunneling of these droplets through a flowing soap film. The energy
barrier $E_{C}(h)$ is found to be linearly proportional to the film
thickness $h$ with the result: $E_{C}=E_{min}(1+\alpha h/R_{0})$.
Here the minimal barrier height $E_{min}(\equiv\frac{1}{2}mv_{C0}^{2})\simeq0.01\,\text{erg}$
and the slope $\alpha=3.9$ are determined. The measured $E_{min}$
corresponds to the creation of an excess surface area of $\sim14\pi R_{0}^{2}$,
which turns out to be consistent with the Rayleigh instability condition
of pinching off at $\lambda_{max}/2$. The observed $E_{min}$ also
implies the existence of a critical Weber number $\text{W\ensuremath{e_{C}}}(\equiv2\rho_{w}R_{0}v_{C0}^{2}/\sigma)\simeq44$,
when the film inertia is unimportant, $h\rightarrow0$. 

A self consistent theory is not currently available, and we wish that
our observations will provide a useful foundation for such theory.
The dynamics is clearly complex in that it involves multiple length
and time scales. We have identified two such scales, $R_{1}$ and
$R_{2}$, that are needed to account for the energy and momentum exchanges
between the droplet and the film. A self consistent theory must deal
with additional length scales, corresponding to early-time or short-length-scale
events, where $\text{Re}$ is small, and vorticity production, hence
energy dissipation, is prominent. The physics in this regime may explain
the intriguing observation that the energy dissipation becomes negligible
for some macroscopically thin but microscopically thick films; i.e.,
the limit $h\rightarrow0$ must correspond to a film that is still
thick enough so that it can be stretched to the Rayleigh instability
limit. Another issue of interest is the separation dynamics of a tunneling
droplet from the rest of the film. This is a singular event that produces
discontinu\textcolor{black}{ities, such as $v_{x}\neq v_{x}'$ at
the moment of separation.}

This investigation was initially motivated by our desire to print
ink patterns in a turbulent flowing soap film and to study how different
spatial modes of the pattern are dispersed by turbulent eddies. This
would allow passive-scalar turbulence to be studied in a controllable
fashion with a defined initial condition. Our measurements presented
above give a parameter range for ink droplets to remain on the surface
of a moving film, which is a prerequisite for a successful conduct
of such measurement. We also found that a ballistic water droplet
is an effective wave generator; we had no difficulty of observing
the anti-symmetric waves in the film and were able to precisely determine
their speeds for different film thicknesses. Curiously, however, the
symmetric waves remain elusive. Unlike anti-symmetric waves, the symmetric
wave is an important attribute of a soap film and would allow experimenters
to obtain useful information about Marangoni elasticity, which is
not readily measured in a film. The failure to observe such a wave
suggests that the peristaltic oscillations must decay fast and hence
not be detectable in our current experimental setting. We wish to
examine this issue more carefully in future experiments.

\textcolor{black}{Aside from its academic interest, the ability of
small particles to penetrate a fluid film without damaging it can
have important technological applications such as encapsulation of
solid particles and transmission of genetic materials through biological
cells. The latter is a fascinating application of ballistic transmission
in biological systems where gold particles coated with DNA molecules
of interest can be be delivered into plant or animal cells \cite{Lin2000}.
A better understanding of transmission kinematics, such as the one
studied here, may shed new light on its working principle and can
ultimately improve the quality of this important technology.}

\section{Acknowledgment}

This work is supported by the NSF under the grant no. DMR-0242284.

\bibliographystyle{apsrev}
\bibliography{droplet2007a}

\end{document}